\documentclass[prd, twocolumn,floats,floatfix,nofootinbib,11pt] {revtex4-1}
\usepackage{graphicx}
\usepackage{dcolumn}
\usepackage{bm}
\usepackage{graphics}
\usepackage{slashed}
\usepackage{amssymb}
\usepackage{natbib}
\usepackage{amsmath}
\usepackage{url}
\usepackage[usenames,dvipsnames,svgnames,table]{xcolor}
\newcommand{\bea}{\begin{eqnarray}}
\newcommand{\ena}{\end{eqnarray}}
\newcommand{\bean}{\begin{eqnarray*}}
\newcommand{\enan}{\end{eqnarray*}}

\begin{document}

\title{Nontrivial Causal Structures Engendered by Knotted Solitons}

\author{
E. Goulart\\ 
CAPES Foundation, Ministry of Education, Bras\'ilia/DF - Brazil\\
and\\
D.A.M.T.P.,\\  University of Cambridge, U.K.\\\texttt{egoulart@cbpf.br}}

\begin{abstract}
It is shown that the causal structure associated to string-like solutions of the Fadeev-Niemi (FN) model is described by an effective metric. Remarkably, the surfaces characterising the causal replacement depend on the energy momentum tensor of the background soliton and carry implicitly a topological invariant $\pi_{3}(\mathbb{S}^2)$. As a consequence, it follows that the pre-image curves in $\mathbb{R}^3$ nontrivialy define directions where the cones remain unchanged. It turns out that these results may be of importance in understanding time dependent solutions (collisions/scatterings) numerically or analytically.

\end{abstract} 

\maketitle

\section{Introduction}
The existence of closed string-like solutions in (3+1)-dimensional field theories is certainly one of the intriguing aspects of modern mathematical physics. Typically, these localised solutions describe one-dimensional structures which may twist non-trivialy in the forms of loops, links and knots characterised by a Hopf index. Roughly, a Hopf soliton (or hopfion) is a knot in a three-dimensional continuous unit vector field which cannot be unknotted without cutting \cite{manton}. Remarkably, they appear in a variety of physical systems such as Bose-Einstein condensates \cite{BE}, ferromagnetism \cite{ferro}, magnetohydrodynamics \cite{magneto} and non-abelian gauge theories, where they are supposed to describe glue-balls \cite{Fad1}. More recently, it was proposed  that liquid crystals also provide an ideal setting for exploring such topological phenomena \cite{nematics}\footnote{See also \cite{Radu} for a discussion in the context of steady Euler flows}.

One of the simplest relativistic systems supporting knots is the $O(3)$ variant of the Skyrme model \cite{skyrme}. The so called Fadeev-Niemi (FN) model describes the dynamics of a three-dimensional iso-vector $\textbf{n}(x^{a})$ taking values on a bi-dimensional sphere $\mathbb{S}^2$ i.e. $\textbf{n}\cdot\textbf{n}=1$ \cite{Fad1}. The lagrangian is that of a sigma model plus fourth order corrections and the topological content appears when we consider static solutions with asymptotic behaviour $\textbf{n}|_{\infty}=(0,0,1)$. In this situation the field realizes the map between spheres $\mathbb{R}^{3}\cup \left\{\infty\right\}\cong\mathbb{S}^{3}\rightarrow \mathbb{S}^{2}$ which is characterised by the Hopf index. Interestingly, the FN model appears quite naturally in the dual superconducting (DS) picture of the strongly coupled SU(2) Yang-Mills theory discussed in a series of papers \cite{Fad2}-\cite{Fad6}. Accordingly,  the high energy limit of the theory describes asymptotically free, massless point-like gluons and the infrared limit describes extended flux tubes which close on themselves in stable knotted configurations. This scenario is particularly compelling because it is consistent with the accepted notion of color confinement in QCD and therefore can shed some light into the mass gap problem. Unfortunately, the equations are highly nonlinear and most of the results rely on numerical lattices approaches and/or approximations \cite{numerical1}-\cite{numerical4}. 

Although important results on the global existence and development of singularities have been obtained for semi-linear wave maps (see, for instance, \cite{Shatah} -\cite{tataru} and references therein) much less is known about the evolutionary properties of the FN equations. Generically, one expects that not all initial data will be mathematically admissible since for a large class of them the Cauchy problem would be ill-posed. Indeed, it is quite common that quasi-linear PDE's generate systems which are not of evolutionary type (for some data) even if the theory is Lorentz invariant by construction (see, for instance, \cite{Er}). Needless to say, this lack of hiperbolicity may be of crucial importance in numerical simulations where imprecisions in the initial data may originate instabilities, singularities and discontinuous solutions.

In this letter we investigate the causal structure of the FN model and clarify some aspects related to its propagation features. We show that the high frequency excitations on top of background solutions are described by characteristic surfaces governed by a curved effective geometry (see \cite{Viss} for a review). This is related to the fact that for quasi-linear equations wave velocities are not given a priori, but change as functions of initial data, directions of propagation and wave polarisation. With respect to the new geometry, rays are equivalent to null geodesics, and therefore can be described using traditional tools of General Relativity (see \cite{christ} for a similar analysis in the context of hydrodynamics).  In particular, we show that the causal replacement inherits the Hopf charge of the soliton. In this sense, the present work is a natural generalisation of previous results sketched by Gibbons and the author in \cite{GWG}.

\section{Fadeev-Niemi Model}

\subsection{Kinematics}

Formally, the FN model is a Lorentz invariant lagrangian theory of maps into a surface \cite{wong}. It can be implemented in terms of a continuous and surjective map  $\phi:(\mathbb{R}^{1+3},\eta)\rightarrow (\mathbb{S}^{2},h)$ where $\eta_{ab}=\mbox{diag}(+---)$ and $h_{AB}(\phi)$ is the riemannian metric on the unit 2-sphere. The map induces the pull-back spacetime tensors $\phi^{*}h$ and $\phi^{*}\epsilon$
\begin{eqnarray*}
L_{ab}:&=& h_{AB}(\phi)\partial_{a}\phi^{A}\partial_{b}\phi^{B}\\
F_{ab}:&=&\epsilon_{AB}(\phi)\partial_{a}\phi^{A}\partial_{b}\phi^{B},
\end{eqnarray*}
with $\epsilon_{AB}$ the area 2-form on the sphere. Following \cite{MAN} we call the tensor $\eta^{-1}\circ \phi^{*}h$ the strain for the map $\phi$. The antisymmetric object $F_{ab}$ and the strain satisfy the algebraic relation
\begin{equation}\label{double}
F^{a}_{\phantom a c}F^{cb}=L^{a}_{\phantom a c}L^{cb}-L^{c}_{\phantom a c}L^{ab}.
\end{equation}

Typically, one is interested in particular mappings with the asymptotic behaviour $\phi^{A}|_{\infty}\rightarrow const$, which means that the state of the fields are homogeneous as they approach spatial infinity\footnote{This is related to the requirement of finiteness of the energy for static configurations.}. As usual, we choose this constant such that the field state corresponds  to the north pole $\textbf{N}$ on the target. Under this boundary condition, $\phi^{A}(x^{0},x^i)$ effectively maps $\mathbb{S}^{3}$ in $\mathbb{S}^2$ for a given time coordinate $x^{0}$ and therefore there is a homotopy invariant $Q=\pi_{3}(S^{2})\in\mathbb{Z}$. Roughly, it will remain the same under any smooth deformation of the map.

According to the above construction $F_{ab}$ is  globally exact i.e.
\begin{equation}
\partial_{a}F_{bc}+\mbox{cyclic}=0.
\end{equation}
Thus, $\exists\ C_{a}$ such that $F_{ab}=\partial_{[a}C_{b]}$ everywhere. Also, if we define the dual $\stackrel{\ast}{F_{ab}}\equiv\eta_{abcd}F^{cd}/2$, with $\eta_{abcd}$ the usual Levi-Civita tensor it follows that
\begin{equation}
\stackrel{\ast}{F_{ab}}F^{ab}=0,
\end{equation}
identically. In particular we have $\partial_{b}\stackrel{\ast}{F^{ab}}=0$ and it results a conserved current of the form $J^{a}=\stackrel{\ast}{F^{a}_{\phantom a b}}C^{b}$. Whitehead first showed \cite{white} that it is possible to express the Hopf invariant as an integral of the form 
\begin{equation}
 Q\propto \int_{\mathbb{R}^3} \eta_{abcd}F^{ab}C^{c}t^{d}dx.
\end{equation}
with $t^{b}$ a normalised timelike vector  ($t^{a}t_{a}=1$) orthogonal to the space slices. Interestingly, the pre-image $\phi^{-1}(P)$ of a given point $P\in\mathbb{S}^{2}$ is an integral line of the divergenceless magnetic-like vector field in $\mathbb{R}^{3}$ given by
\begin{equation}
B^{a}\equiv \stackrel{\ast}{F^{a}_{\phantom ab}}t^{b}.
\end{equation}
As is well known, $Q$ can be interpreted heuristically as the linking number between two such ``magnetic" field lines, ($\phi^{-1}(P_{1})$ and $\phi^{-1}(P_{2})$, respectively). Generically, these field lines will twist among themselves in a highly nontrivial way for large values of the index, filling the space with a complex fibered structure (see Fig.1). In particular, the pre-image of the south pole $\textbf{S}$ is defined as the position of the Hopf soliton as it corresponds to the position in $\mathbb{S}^{2}$ more distant to the vacuum.

\begin{figure}[h]
       \centering  
       \includegraphics[scale=.5]{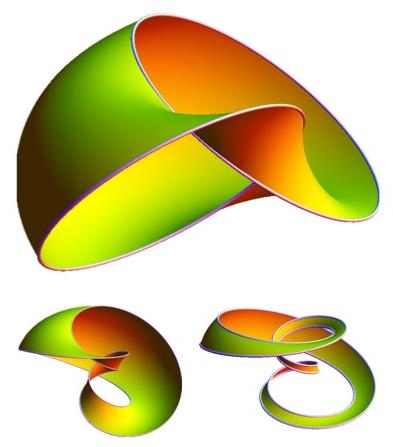}
       \caption{Pencils of ``magnetic" field lines in $\mathbb{R}^{3}$ for $N=1,2\ \mbox{and}\ 3$, respectively. The closed curves representing the boundaries of the (Seifert) surfaces correspond to the pre-images $\phi^{-1}(P_{1})$ and $\phi^{-1}(P_{2})$ of two points $P_{1}$ and $P_{2}$ in $\mathbb{S}^{2}$.}
       \label{dynamic}
\end{figure}

\subsection{Dynamics}
The pulled-back tensors $L_{ab}$ and $F_{ab}$ give rise to invariants from which we can naturally define an action for the map. In the model proposed by Fadeev and Niemi the action is given by
\begin{equation}\label{1}
S[\phi]=\int \ \frac{1}{2}L^{a}_{\phantom a a}-\frac{\kappa^{2}}{4}F_{ab}F^{ab} \ d^{4}x,
\end{equation}
where, as usually, the second term is introduced in order to guarantee the stability of solitons against scalings and $\kappa$ is a parameter controlling the strength of nonlinearities. We note that this term resembles the lagrangian of Maxwell's electrodynamics\footnote{When one considers static configurations, the first term corresponds to a Dirichlet term and the second to the energy of the vector field $B^{a}$ directed along the fibers \cite{slobo1}.}. 
 
Variation with respect to $\phi^{A}$ yields a system of second order quasi-linear PDE's which can be written in the compact form
\begin{equation}\label{1}
\left(H_{AB}\partial^{a}\phi^{B}\right)_{||C}\partial_{a}\phi^{C}=0,
\end{equation}
where 
\begin{equation}
H_{AB}\equiv h_{AB}- \kappa^{2}\epsilon_{AP}\epsilon_{BQ}X^{PQ},
\end{equation}
$X^{AB}\equiv\partial_{a}\phi^{A}\partial_{b}\phi^{B}\eta^{ab}$ and $||$ represents the covariant derivative with respect to $h_{AB}$ \footnote{Note that Eq. (\ref{1}) can be written also in the form $[(h_{AB}\eta^{ab}-\kappa^{2}\epsilon_{AB}F^{ab})\partial_{b}\phi^{B}]_{||C}\partial_{a}\phi^{C}=0$}. In terms of the target connection $\Gamma^{A}_{\phantom a BC}$, Eq. (\ref{1}) can be written as
\begin{equation}\label{bse}
\partial_{a}\left(H_{AB}\partial^{a}\phi^{B}\right)=\Gamma^{D}_{\phantom a AC}H_{DB}\partial^{a}\phi^{B}\partial_{a}\phi^{C},
\end{equation}
which reveals that the equation of motion consists of various types of self-interactions arising from the non-standard kinetic terms and the target space geometry.  

Generically, it is possible to express Eq. (\ref{bse}) as
\begin{equation}\label{L}
M^{ab}_{\phantom a\phantom a  AB}(\phi,\partial\phi)\ \partial_{a}\partial_{b}\phi^{B}+...=0,
\end{equation}
where $``..."$ stands for semilinear terms in $\phi^{A}$ (lower order derivatives) and the principal symbol is given by
\begin{equation}\label{M}
\textbf{M}=\eta^{ab}H_{AB}+\frac{\kappa^{2}}{2} \epsilon_{AP}\epsilon_{BQ}\partial^{a}\phi^{(Q}\partial^{b}\phi^{P)}.
\end{equation}

As it is well known, the highest-order terms in derivatives almost completely controls the qualitative behaviour of solutions of a partial differential equation. We note that $\textbf{M}$ is automatically symmetric with respect to $ab$ and $AB$. Also, in the absence of the Skyrme term eq. (\ref{1}) reduces to the semilinear equation known as the classical $O(3)$ sigma-model.  
 
In what follows we will see that the energy momentum tensor plays a crucial role in the description of the propagation cones. Here, $T^{ab}$ can be split in two parts  
\begin{eqnarray*}
T^{ab}_{(1)}&=&L^{ab}-L^{c}_{\phantom a c}g^{ab}/2,\\
T^{ab}_{(2)}&=&[F^{a}_{\phantom a c}F^{cb}+F_{cd}F^{cd}g^{ab}/4]\kappa^2.
\end{eqnarray*}
Note that $T^{ab}_{(2)}$ has exactly the same form as the energy momentum-tensor in electrodynamics. In particular, it is traceless, which implies the well known equation of state $p_{(2)}=\rho_{(2)}/3$ between the corresponding energy and pressure. In \cite{gib} it was showed by Gibbons that the energy momentum tensor of the $SU(2)$ Skyrme model satisfies the dominant energy condition. This was achieved using the eigenvalue decomposition introduced by Manton in \cite{MAN}. Here, we see that this property still holds because the content of the lagrangian is the same.  

\section{Characteristic Surfaces}

The nonlinear structure of the field equations implies that linearised waves interact with background solutions in a nontrivial way. The characteritic surfaces of the model can be obtained by means of the eikonal approximation.  Formally, we consider a one-parameter family of solutions of the form
\begin{equation}
\phi^{A}(x)=\phi^{A}_{0}(x)+\alpha\varphi^{A}(x)exp\left(i\Sigma(x)/\alpha\right),
\end{equation}
where $\phi^{A}_{0}(x)$ is a smooth solution and let the real parameter $\alpha\rightarrow 0$. In this limit, we can discard all semilinear contributions in (\ref{L}) and consider only the principal part term contributions. 

Defining the wave covector $k_{a}\equiv \partial_{a}\Sigma$, the equation of motion reduces to the eigenvalue equation
\begin{equation}\label{eigenvalue}
\left[M_{AB}(\phi_{0}, k)\right]\varphi^{B}=0,
\end{equation}
where we defined the symmetric matrix $M_{AB}(\phi_{0},k)\equiv M^{ab}_{\phantom a\phantom a AB}(\phi_{0})k_{a}k_{b}$. It follows that (\ref{eigenvalue}) can be solved only if $k_{a}$ satisfy the algebraic conditions
\begin{equation}
F_{x}(\phi_{0},k)\equiv det(M_{AB}(\phi_{0},k))=0.
\end{equation}
As a consequence, at a given spacetime point, the wave normals are characterised by the roots of a multivariate polynomial of fourth order in $k_{a}$ in the cotangent space $T_{p}^{*}$. The resulting algebraic variety changes from point to point in a way completely prescribed by the background solution $\phi^{A}_{0}$ and the nonlinearities of the model.  

The general form of $F_{x}$ is given by a quartic polynomial which factorizes. In other words the characteristic polynomial reduces to a product of two simpler quadratic terms satisfying
\begin{equation}\label{z}
[\eta^{ab}k_{a}k_{b}][(h^{-1})^{cd}k_{c}k_{d}]=0.
\end{equation}
Surprisingly, the reciprocal quadratic form $(h^{-1})^{ab}$ can be written in terms of the total energy momentum tensor of the background field
\begin{equation}\label{E}
(h^{-1})^{ab}\equiv(1-\kappa^{2}\mathcal{L})\eta^{ab}+\kappa^{2}T^{ab},
\end{equation}
where $\mathcal{L}$ is nothing but the lagrangian of the model and $T^{ab}=T^{ab}_{(1)}+T^{ab}_{(2)}$. As a consequence, the vanishing sets of (\ref{z}) constitute the FN analogues of the Fresnel equation encountered in optics. They play the role of a fourth order space-time dispersion relation (at least up to a conformal factor).  In general, $\Sigma(x)$ will solve one quadratic polynomial or the other, although it is possible that there exist some directions where the vanishing sets coincide. Consequently, the model admits two different types of waves. One wave travels with the velocity of light while the other travels with a velocity wich depends implicitly on the solution and on the Hopf charge. More explicitly, we have
\begin{equation}\label{quantity}
(h^{-1})^{ab}=(1-2\kappa^2 \mathcal L)\eta^{ab}+\kappa^{2}(L^{ab}+\kappa^{2}F^{a}_{\phantom a c}F^{cb})
\end{equation}
which reveals that both pulled-back tensors contribute to the causal structure. Note that if $\kappa$ is set to zero the FN model reduces to the usual $O(3)$ sigma model (wave map equation) implying that the effective metric becomes flat everywhere.

Now, if the quantity $(h^{-1})^{ab}$ is non-degenerate it is possible to define its inverse $h_{ab}$ such that $(h^{-1})^{ac}h_{cb}=\delta^{a}_{\phantom a b}$.  In general, the \textit{effective metric} $h_{ab}$ defines a Lorentzian metric on spacetime, the null cones of which are the effective ``sound cones" of the theory. The ray vectors $q^{a}$ associated to the wave fronts are the vanishing sets of the dual polynomial $G_{x}$
\begin{equation}\label{w}
G_{x}(\phi_{0},q)\equiv [\eta_{ab}q^{a}q^{b}][h_{ab}q^{a}q^{b}]=0
\end{equation}
As is well known, these cones completely determine the causal structure of the theory once a solution is given.  In particular, nontrivial excitations propagate along geodesics of the effective spacetime. In fact, we obtain (see, for instance, \cite{ngeo})
\begin{equation} 
(h^{-1})^{ab}k_{c;a}k_{b}=0
\end{equation}
where ; is such that ${h}_{ab;c}=0$.  Note that for an arbitrary solution $\phi_{0}$ endowed with invariant $Q$ the quadratic form $h$ is generally curved, implying that we can use appropriate geometrical methods to describe the causal structure.   

\section{Qualitative analysis}

We now discuss some general properties of the effective metric. For the sake of simplicity, let us suppose that the solution $\phi^{A}_{0}$ describes a static Hopf soliton with a given topological invariant $Q$. Time-dependent configurations may be easily obtained using the same framework. In the static regime Eq. (\ref{quantity}) reduces to
\[(h^{-1})^{ab}(\vec{r})= \left( \begin{array}{cc}
(h^{-1})^{00} & 0  \\
0 & (h^{-1})^{ij}  \\
 \end{array} \right)\] 
with $\vec{r}$ the position vector in $\mathbb{R}^{3}$ and $i,j=1,2,3$. Because the field tends to its vacuum in spatial infinity i.e. $\phi^{A}|\infty\rightarrow const$, we automatically have $(h^{-1})^{ab}|_{\infty}\rightarrow \eta^{ab}$ which means that the effective geometry is static and asymptotically flat for all possible static solitons. This is an expected result since the solitons are supposed to be localised field structures in space with finite energy.

We now ask the following: locally, how does a wave propagate in the direction of $B^{a}$? In other words, what is the velocity of a disturbance which evolves in the direction defined by a pre-image curve at a spacetime point $p$? In order to find this velocity, first note that the vector $B^{a}$ is automatically an eigenvector of the linear operator $L^{a}_{\phantom a b}$ with a null eigenvalue. Indeed, in the static regime we have 
\begin{equation}
L^{a}_{\phantom a b}t^{b}=0,\quad\quad\quad\quad L^{a}_{\phantom a b}B^{b}=0,
\end{equation}
meaning that the kernel of $L^{a}_{\phantom a b}$ is determined by a timelike vector $t^{a}$ and the ``magnetic" field itself. A direct inspection in Eq. (\ref{double}) implies that the projection $F^{a}_{\phantom a c}F^{c}_{\phantom a b}B^{b}$ also vanishes. Now, the wave front is determined by a covector $k_{a}\in T_{p}^{*}M$ of the form $k_{a}=rt_{a}+sB_{a}$ with (r,s) real components.   Using the above results in Eq. (\ref{quantity}) it follows directly
\begin{equation}
(h^{-1})^{ab}k_{a}k_{b}=0\quad\rightarrow \quad\eta^{ab}k_{a}k_{b}=0.
\end{equation}
Thus, the vector field $B^{a}$ defines special directions where the effective cone coincides with the Minkowski cone. In other words, the pre-images of the map $\phi^{A}$ in $\mathbb{R}^3$ define directions where the effective causal structure is not affected by the solution. Note however that, because the pre-image curves are linked in a nontrivial way, the resulting light cones in the global may have a very complicated structure. 

\begin{figure}[h]
       \centering  
       \includegraphics[scale=.4]{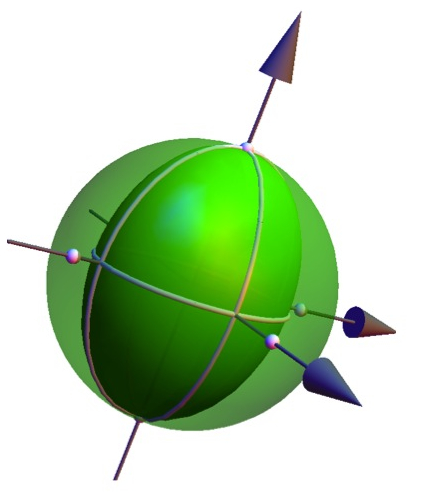}
       \caption{Cross-section of the effective ``light" cone in $T_{p}M$. The round sphere represents velocities of propagation according with the Minkowski metric. Note that they intersect in the direction of $B^{a}$.}
       \label{dynamic}
\end{figure} 
What about wave propagation in other directions? Following Manton \cite{MAN} we suppose that $L_{ab}$ can be diagonalised relative to $\eta_{ab}$ in a given spacetime point $p$. As is well known the eigenvalues are necessarily nonnegative and due to rank considerations two of them vanish identically. Choosing the spatial axis in such a way that $B_{a}=(0,0,0,B_{3})$ it follows
\begin{eqnarray*}\label{eigenv}
&&L_{ab}=\mbox{diag}(0,\lambda_{1}^{2},\lambda^{2}_{2},0)\\
&&F^{a}_{\phantom a c}F^{cb}=\mbox{diag}(0,\lambda_{1}^{2}\lambda^{2}_{2},\lambda_{1}^{2}\lambda^{2}_{2},0).
\end{eqnarray*}
In particular, we have $B^{a}B_{a}=-\lambda_{1}^{2}\lambda_{2}^2$. Using Eq. (\ref{eigenv}) in Eq. (\ref{quantity}) it results an automatically diagonal $(h^{-1})^{ab}$
\begin{eqnarray*}
(h^{-1})^{00}&=&+(1+\kappa^2\lambda^{2}_{1})(1+\kappa^2\lambda^{2}_{2})\\
(h^{-1})^{11}&=&-(1+\kappa^2\lambda^{2}_{2})\\
(h^{-1})^{22}&=&-(1+\kappa^2\lambda^{2}_{1})\\
(h^{-1})^{33}&=&-(1+\kappa^2\lambda^{2}_{1})(1+\kappa^2\lambda^{2}_{2}).
\end{eqnarray*}
Two important results emerge: i) The effective metric has a Lorentzian signature $(+---)$ for all possible static solitons \footnote{At least if the quantities $\lambda_{i}$ are sufficiently regular everywhere.}. This means that the linearisation of the equations of motion (\ref{bse}) on top of a static solution yields a well defined causal structure. ii) Wave propagation is always sub-luminal, implying that the theory is causal. In particular, velocities of propagation orthogonal to the pre-image curve are of the form $c_{i}^{2}=(1+\kappa^{2}\lambda_{i}^2)^{-1}$ with $i=1,2$. Note that they tend to $1$ in spatial infinity.  
 
The ``sound" cones of the theory may be obtained also in terms of the null intervals
\begin{eqnarray}\label{dist}
(dx_{0})^2&-&(1+\kappa^{2}\lambda_{1}^{2})(dx_{1})^{2}-\\
&-&(1+\kappa^{2}\lambda_{2}^{2})(dx_{2})^{2}-dx_{3}^2=0.\nonumber
\end{eqnarray}
It follows that, locally, the cross-sections of the effective cone in the tangent space are described by ellipsoids with major-axis oriented in the directions defined by $B^{a}$. This result is valid for all possible pre-images $\phi^A=const$ for all possible charges. See Fig.2 for an illustration. Again, we stress that because links between pre-images are nontrivial and eigenvalues are space-dependent quantities there will be a highly complex causal structure associated. It remains the possibility that linked or knotted congruences of rays emerge in $\mathbb{R}^{3}$ with regions where waves are trapped.

\section{Rational Maps and approximate Effective Geometries}

In this section we use the rational map \textit{ansatz} introduced by Suttcliffe in \cite{numerical4} (see also \cite{isos}) in order to explore approximate expressions for $(h^{-1})^{ab}$. The idea is to achieve a qualitative picture of the metrics without appealing to complicated numerical simulations. 

The strategy is as follows. Recall that the assumption $\phi^{A}|_{\infty}=const$ effectively compactifies $\mathbb{R}^3$ to the hypersphere $\mathbb{S}^{3}$.  A point of $\mathbb{S}^{3}$ can be thought also as a point of the plane of complex dimension 2 with coordinates $(Z_{1},Z_{0})$ and $|Z_{1}|^{2}+|Z_{0}|^{2}=1$. Project the point $(Z_{1},Z_{0})$ onto $\mathbb{R}^{3}$ using the map 
\begin{equation}
(Z_{1},Z_{0})=\left(\frac{x_{1}+ix_{2}}{r}sinf,cosf+i\frac{sinf}{r}x_{3}\right),
\end{equation}
where $(x_{1},x_{2},x_{3})$ are cartesian coordinates, $r=(x_{1}^2+x_{2}^2+x_{3}^2)^{1/2}$ and $f(r)$ is a monotonically decreasing function such that $f(0)=\pi$ and $f(\infty)=0$. Interpret $\mathbb{S}^2$ as a Riemann sphere introducing a complex coordinate $W$. The idea is to start with spherical coordinates $\phi^{A}=(\Theta,\Phi)$ and metric $h_{AB}=\mbox{diag}(1,\mbox{sin}^{2}\Theta)$. Then, perform a stereographic projection from the south pole $\textbf{S}$ ($\Theta=\pi$) to the equatorial plane to obtain $W=R e^{i\Phi}$ with the absolute value given by $R=tg(\Theta/2)$. Finally, write $W$ as a rational function of the complex quantities $Z_{1}$ and $Z_{0}$ in the form
\begin{equation}\label{mapping}
W=p(Z_{1},Z_{0})/q(Z_{1},Z_{0})
\end{equation}
where $p$ and $q$ are polynomials. 

As a consequence, for each point $W\in\mathbb{S}^2$, we obtain a closed curve $\phi^{-1}(W)$ in $\mathbb{R}^3$, whose image by $\phi^{A}$ is the point $W$\footnote{It is also common to call the pre-image as the ``fiber" over the point $W$.}. The tangent vectors to these curves define precisely the direction of $B^{a}$. In particular, we can obtain the whole of the $\mathbb{R}^{3}$ as ``fibers" over the ordinary 2-sphere. Incidentally, for any two points, $W_{1}$ and $W_{2}$ on $\mathbb{S}^{2}$, the corresponding ``fibers" are nontrivialy linked. The number of links depends crucially on the degree of the polynomials $p(Z_{1},Z_{0})$ and $q(Z_{1},Z_{0})$.

The interesting point here is that once $p$ and $q$ are given as functions of $Z_{1}$ and $Z_{0}$, the effective metric may be easily obtained. This can be accomplished writing the pull-back quantities $L_{ab}$ and $F_{ab}$ in terms of the derivatives of $W$. We obtain
\begin{eqnarray}
&& L_{ab}=\frac{2}{(1+|W|^{2})^{2}}\partial_{(a}W\partial_{b)}\overline{W},\\
&& F_{ab}=\frac{2i}{(1+|W|^{2})^2}\partial_{[a}W\partial_{b]}\overline{W}.
\end{eqnarray}
with $(a,b)=(ab+ba)$ and $[a,b]=ab-ba$ and $\overline{W}$ the complex conjugate\footnote{ Alternatively, we have:
\begin{eqnarray*}
&& L_{ij}=\frac{4}{(1+R^2)^2}\left(\partial_{i}R\partial_{j}R+R^2\partial_{i}\Phi\partial_{j}\Phi \right)\\
&& F_{ij}=\frac{4}{(1+R^2)^2}\partial_{[i}R\partial_{j]}\Phi 
\end{eqnarray*}}. Basically, there will be three different classes of effective geometries. They follow naturally from the classification presented in \cite{numerical4} for all possible background fields. Roughly, they describe the causal structures associated to:
\begin{enumerate}
\item{Toroidal fields $\mathcal{A}_{n,m}$: $W=Z_{1}^{n}/Z_{0}^{m}$, with $n,m\in \mathbb{Z}$ and N=n.m;}
\item{ (a,b)-torus knots $\mathcal{K}_{ab}$:  $W=Z_{1}^{\alpha}Z_{0}^{\beta}/(Z_{1}^{a}+Z_{1}^{b})$, with $\alpha$ a positive integer, $\beta$ a non-negative integer and a,b co-prime positive integers with $a>b$ and $N=\alpha b+\beta a$;}
\item{Linked Hopfions $\mathcal{L}^{\alpha,\beta}_{p,q}$: $W=Z_{1}^{\alpha}Z_{0}^{\beta}/(Z_{1}^{p}+Z_{1}^{q})$ with $\alpha$ a positive integer, $\beta$ a non-negative integer $p,q$ not co-prime;}
\end{enumerate}

As a consequence of the above \textit{ansatz}, one is now able to speak about effective geometries engendered by knotted structures. Each geometry carries implicitly a Hopf charge with it and is somehow related to the homotopy class of the configuration. Waves are described by pencils of null geodesics scattered by the corresponding geometry and it is possible that they remain trapped for some regions of the effective spacetime. In order to understand better these aspects, it would be interesting to analyse in more details the geometrical and topological properties of the above metrics, including their symmetries (see \cite{gib2} for a Bianchi classification in the helical phase of chiral nematic liquid crystals). In particular, one could investigate the approximate behaviour of geodesics arriving from a vacuum domain which is scattered by the solitonic structure. We shall come back to these questions in the future. 
\subsection{Example: Q=1 Hopfion}
As a final remark, lets us illustrate the qualitative behaviour of the pre-images in $\mathbb{R}^{3}$ for the $Q=1$ Hopfion. As we discussed above they define directions where the waves ``perceive" the actual Minkowski spacetime. In other words, metrical relations are only distorted for intervals having some component orthogonal to $B^{a}$.

In this case the rational map is simply $W=Z_{1}/Z_{0}$. For the sake of simplicity we choose a profile function with an exponential decay of the form $f(r)\equiv \pi e^{-r^2}$. We can now calculate the absolute value (R) and the argument ($\Phi$) as explicit functions of the coordinates $(x_{1},x_{2},x_{3})$. Now, instead of only considering pre-images of points, we will consider pre-images of whole sets on $\mathbb{S}^2$. In particular, we consider the pre-images of the paralels $R=const$ and the meridians $\Phi=const$. They characterise two families of surfaces in $\mathbb{R}^{3}$.  Surfaces with constant $R$ are homeomorphic to tori. The tori are nested, and their size increase as $R$ decreases. In particular, one obtains a one-dimensional curve in the limit $R\rightarrow\infty$. This curve is precisely the position of the Hopfion. Surfaces with constant $\Phi$ are not so simple. They are homeomorphic to parabolic Dupin cyclides, i.e. a specific inversion of the torus.
\begin{figure}[h]
       \centering  
       \includegraphics[scale=.4]{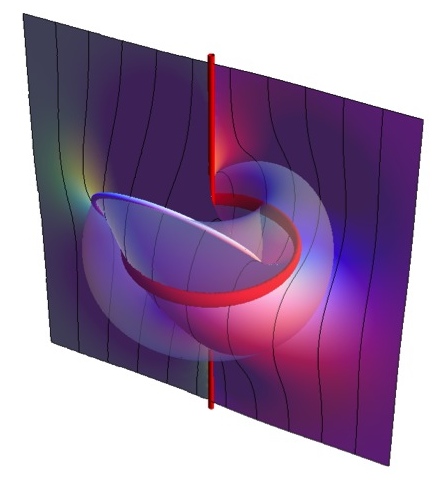}
       \caption{Intersection of a torus and a parabolic Dupin cyclide. The resulting closed curve is the pre-image of a point in the unit sphere where a parallel meets a meridian. The figure shows also the position of the Hopfion and the pre-image of infinity. }
       \label{cyclide}
\end{figure} 

The points where a parallel intersects a meridian in $\mathbb{S}^{2}$ are given by the intersections of $\phi^{-1}(R=const)$ and $\phi^{-1}(\Phi=const)$ in $\mathbb{R}^{3}$. They are closed curves pairwise linked. We can see such a general behaviour in (FIG. 3). The figure shows the intersection between the meridian $\Phi=0$ and a parallel $R=1$. Now, if we consider the collection of all points lying in $\mathbb{S}^{2}$ we obtain the congruence of ``magnetic" field lines filling the entire space (see FIG. 4). They can be thought as a continuous deformation of the usual Hopf fibration \cite{hopfib}. Roughly, effective metrical relations are the same in the directions of the fibres while they are distorted in any other direction, leading to a nontrivial causal structure in the global.

\begin{figure}[h]
       \centering  
       \includegraphics[scale=.4]{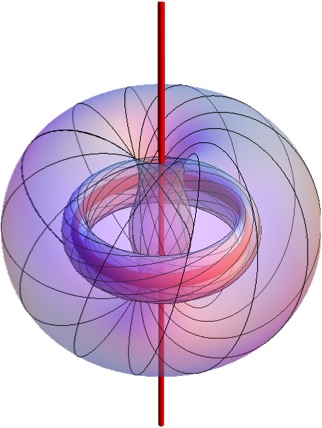}
       \caption{Global behavior of the ``magnetic" field lines. Metrical relations remain the same in the directions of the fibers while they are distorted in any other direction.}
       \label{dynamic}
\end{figure}

\section{Acknoweledgement} E. Goulart would like to thank CAPES - Brazil proc. 2383136 for financial support and G. W. Gibbons for comments and suggestions.

\end{document}